\documentclass[prl,twocolumn,twoside,preprintnumbers,superscriptaddress]{revtex4}
\usepackage{color}

\usepackage{setspace}
\usepackage{enumitem}
\usepackage{amsmath}
\usepackage{graphicx}
\usepackage{multirow}
\usepackage{amssymb}
\usepackage{xspace}

\newcommand{\runDM}{\textsc{runDM}\xspace}

\newcommand{\Eq}[1]{Eq.~\eqref{#1}}

\newcommand{\ZZ}{\mathbb{Z}}

\newcommand{\be}{\begin{equation}}
\newcommand{\ee}{\end{equation}}
\newcommand{\bea}{\begin{eqnarray}}
\newcommand{\eea}{\end{eqnarray}}

\begin{document}

\preprint{\vbox{\hbox{SCIPP 17/05}}}

\title{Surprises from Complete Vector Portal Theories:\\ New Insights into the Dark Sector and its Interplay with Higgs Physics}

\author{Yanou~Cui}
\affiliation{Department of Physics and Astronomy, University of California-Riverside, 900 University Ave., Riverside, CA 92521, USA}
\author{Francesco~D'Eramo}
\affiliation{Department of Physics, 1156 High St., University of California Santa Cruz, Santa Cruz, CA 95064, USA}
\affiliation{Santa Cruz Institute for Particle Physics, 1156 High St., Santa Cruz, CA 95064, USA}
\affiliation{Dipartimento di Fisica e Astronomia ``Galileo Galilei'', Universit\`a di Padova, Via Marzolo 8, 35131 Padova, Italy}
\affiliation{INFN, Sezione di Padova. Via Marzolo 8, 35131 Padova, Italy}

\date{\today}
\begin{abstract}
We study UV complete theories where the Standard Model (SM) gauge group is extended with a new abelian $U(1)'$, and the field content is augmented by an arbitrary number of scalar and fermion SM singlets, potentially including dark matter (DM) candidates. Considerations such as classical and quantum gauge invariance of the full theory and S-matrix unitarity, not applicable within a simplified model approach, are shown to have significant phenomenological consequences. The lack of gauge anomalies leads to compact relations among the $U(1)'$ fermion charges, and puts a lower bound on the number of dark fermions. Contrary to naive expectations, the DM annihilation to $Z h$ is found to be p-wave suppressed, as hinted by perturbative unitarity of S-matrix, with dramatic implications for DM thermal relic density and indirect searches. Within this framework, the interplay between dark matter, new vector boson and Higgs physics is rather natural and generic. 
 \end{abstract}
%

\maketitle

\noindent 
{\bf Introduction.} Extra abelian gauge symmetries are among the best motivated extensions to the Standard Model (SM) of particle physics ~\cite{Langacker:2008yv}. Spontaneous breaking of such a $U(1)'$ symmetry is associated with a massive gauge boson $Z'$ that mediates a new type of interaction among SM fields. This $Z'$ boson could also provide a portal to the dark matter sector -- another robust motivation for physics beyond the SM~\cite{Ade:2015xua, Bertone:2004pz}. Extensive studies have pursued this scenario, with simplified models as a commonly employed tool~\cite{Petriello:2008pu,Dudas:2009uq,An:2012va,Frandsen:2012rk,Profumo:2013sca,Buchmueller:2013dya,Alves:2013tqa,Arcadi:2013qia,Lebedev:2014bba,Bell:2014tta,Buchmueller:2014yoa,Harris:2014hga,Alves:2015pea,Chala:2015ama,Alves:2015dya,Alves:2015mua,Fairbairn:2016iuf,Alves:2016cqf,DEramo:2017zqw,Arcadi:2017kky}. Although this approach is advantageous as it allows to study phenomenology with a handful of masses and couplings, some key issues may be missed unless a UV-complete theory is specified.
 
In this paper, we explore the UV-completeness of vector portal models, in particular the implications of the following important theoretical constraints:
\begin{enumerate}[nolistsep,leftmargin=*]
\item Classical level gauge invariance of the theory, including the $U(1)'$ invariance of SM Yukawa terms.
\item Quantum level gauge invariance of the theory, namely the absence of gauge anomalies.
\item Perturbative unitarity of the $S$-matrix.
\end{enumerate}
\textit{These issues, not apparent in a simplified model approach, have profound phenomenological consequences. }

We focus on $U(1)'$ theories where all the new matter fields are SM gauge singlets, and consider an \textit{arbitrary number} of them. The gauge invariance of the SM Yukawa interactions implies that the SM Higgs doublet $H$ typically carries $U(1)'$ charge, unless SM fermions are vector-like under $U(1)'$~\cite{Kahlhoefer:2015bea,Jacques:2016dqz}. This in turn implies a deep connection between DM searches and Higgs physics. 

The cancellation of gauge anomalies is highly non-trivial in a generic $U(1)'$ model. Despite the many constraints, we find very compact relations among the dark gauge charges of the new fermions as well as the SM fields, as provided in Eqs.~\eqref{eq:generalSMsinglets} and \eqref{eq:SMchargesSOLlong}. In particular, we learn from \Eq{eq:generalSMsinglets} that a consistent $U(1)'$ model with new SM singlets require \textit{at least two} species of dark fermions. For DM to play a crucial role in anomaly cancelation, we need \textit{at least three} fermions in chiral representations. 

Perturbative unitarity can impose additional bounds on the model parameters of such theories~\cite{Kahlhoefer:2015bea,Englert:2016joy,Duerr:2016tmh}. We consider for the first time the constraints on the DM annihilation to $Z h$ final state, and derive the bound in \Eq{eq:unitaritybound}. Moreover, we show how unitarity provides the guidelines for necessary processes to be considered. This leads to the realization that the DM annihilation to $Zh$ is p-wave suppressed, with substantial consequences for DM relic abundance and indirect searches. 

Our analysis applies to $U(1)'$ gauge boson and dark sector fields of generic masses. In the last part of this work, we introduce a benchmark model for weak scale $Z'$ and DM, and sketch the phenomenological aspects of this model, emphasizing the interplay with Higgs physics.

\vspace{0.1cm}
\noindent 
{\bf UV-complete theories with a new $U(1)'$ and DM.} We extend the SM matter content by an arbitrary number of fermions $\chi_i$ and scalars $\phi_j$, SM singlets and with dark hypercharges $d_{\chi_i}$ and $d_{\phi_j}$. We employ a notation where all the SM fermions are left-handed Weyl spinors. Quark fields $Q_i$ ($u_i^c$ and $d_i^c$) are doublets (singlets) under the weak-isospin gauge group. Likewise, we introduce lepton doublets $L_i$ and singlets $e_i^c$. The index $i$ runs over the three SM generations. 
SM fields also carry $U(1)'$ charges, denoted by e.g. $d_Q, d_H$. We only consider flavor universal $U(1)'$ gauge charges. The lightest dark fermion $\chi_1 \equiv \chi$ serves as a DM candidate. The DM stability is ensured by proper $U(1)'$ charge assignments to forbid renormalizable couplings such as $L_i H \chi$, or an additional $\ZZ_2$ symmetry when necessary. Dark scalars $\phi_i$ are responsible for the spontaneous breaking of $U(1)'$ by their vacuum expectation value (VEV).

The $U(1)'$ charges are not arbitrary. The gauge invariance of the SM Yukawa terms imposes $d_Q + d_d = - (d_Q + d_u)$ and $d_Q + d_d  = d_L + d_e$. Here, we are assuming that the new gauge symmetry is irrelevant for generating the SM flavor structure. Consequently, the dark hypercharge of the Higgs doublet is  $d_H = d_Q + d_d$, and $H$ is $U(1)'$-neutral only if SM fermions are in a $U(1)'$-vector-like representation.
\newline
 \indent The singlet-only scenario that we focus on is the least constrained given the absence of SM charged exotics. The kinetic mixing~\cite{Holdom:1985ag} between the two Abelian factors is not phenomenologically relevant. This is because the kinetic mixing is suppressed by a loop factor, whereas the $U(1)'$ interaction connects SM and dark fermions at tree-level. Likewise, we assume Higgs portal operators $H^\dag H \phi_i^\dag \phi_j$ to be subdominant. Their impact was studied in Refs.~\cite{Bell:2016fqf,Bell:2016uhg}. The $U(1)'$ gauge boson $V_\mu$ has mass mixing with SM  from $\left| D_\mu H \right|^2 \supset - g_w  \, g_d \, d_H \, v^2_H \,  Z^{{\rm (SM)}}_\mu V^\mu$, where $g_w \equiv \sqrt{g_Y^2 + g_2^2}$, $g_d$ is the $U(1)'$ gauge coupling and the Higgs VEV is $\langle H \rangle^T = (0 \; v_H)$. The mass eigenstates $Z,~Z^\prime$ read
\be
\left( \begin{array}{c} Z^{\rm (SM)}_\mu \\ V_\mu \end{array} \right)  = 
\left( \begin{array}{cc}  \cos\theta & \sin\theta  \\
- \sin\theta &  \cos\theta \end{array} \right)   \left( \begin{array}{c} Z_\mu \\ Z^\prime_\mu \end{array} \right)  \ .
\label{eq:rotation}
\ee
where $Z^{{\rm (SM)}}_\mu$ is the linear combination of neutral $SU(2)_L$ and $U(1)_Y$ gauge bosons corresponding to the SM $Z$ boson. The mixing angle $\theta$ parameterizing the orthogonal rotation can be expressed analytically as
\be
\tan 2 \theta = - \frac{4 g_w\, g_d \, d_H \, v^2_H}{4 g_d^2 (d_H^2 v_H^2 + v^2_\phi) -  g_w^2 v^2_H} \ ,
\label{eq:mixingangle}
\ee
where we define the effective dark VEV as $v_\phi^2 \equiv \sum_i d^2_{\phi_i} v^2_{\phi_i}$. The ElectroWeak Precision Tests (EWPT)~\cite{Erler:2009jh} require that $\theta\lesssim10^{-3}$.
\vspace{0.1cm}\newline
\noindent 
{\bf Anomaly Cancellation.} Even if the classical gauge invariance is satisfied, the theory can be anomalous and additional fermionic degrees of freedom are expected~\cite{Preskill:1990fr}. Despite the arbitrariness of the dark sector, the anomaly condition we find is \textit{extremely compact}.

We require that all the following anomalies vanish: four purely abelian ($U(1)_i U(1)_j U(1)_k = 0$, with $i,j,k = Y, d; U(1)_d\equiv U(1)'$), four mixed ($SU(N)^2 U(1)_i = 0$, with $N = 2, 3$) and two gravitational ($U(1)_i = 0$). The other conditions with SM gauge groups only are automatically satisfied. The dark charges of the new fermions must satisfy the following simple relation 
\be
\sum_{i=1}^n  \left(d_{\chi_i}\right)^3 = \frac{1}{9} \left(  \sum_{i=1}^n  d_{\chi_i} \right)^3 \ .
\label{eq:generalSMsinglets}
\ee
This equation is valid for an arbitrary number $n$ of dark sector fermions. It was also found in Ref.~\cite{Appelquist:2002mw} in the context of neutrino masses in $U(1)'$ models. These models of neutrino masses were also studied in Refs.~\cite{Okada:2016tci,Okada:2016gsh}.

The first noteworthy fact of our result: for $n=1$, the only solution is $d_{\chi}=0$. Thus a consistent vector portal DM theory needs at least two dark fermions! If we add just two fermions, the only solution is a vector-like representation ($d_{\chi_1} = - d_{\chi_2}$), significantly constrained by DM direct searches due to a vector current with the $Z'$. In order to have an axial-vector coupling with the $Z'$, we need at least $n=3$. Interestingly, this is also the first case of a chiral representation, where dark fermions assist anomaly cancellation. With these in mind, for the rest of the paper we focus on considering new chiral fermions. They get \textit{Majorana} mass terms after $U(1)'$ breaking, thus they have a pure axial-vector couplings to $Z'$.  It is desirable to embed the $U(1)'$ into a simple group, in order to avoid Landau poles and facilitate gauge unification~\cite{Slansky:1981yr, Batra:2005rh, Langacker:2008yv}. Since the dark charges in this case must be rational, and abelian gauge charges are always defined up to an overall normalization factor, \Eq{eq:generalSMsinglets} is a cubic Diophantine equation to solve. We list the solutions (up to permutations) with $\min\{|d_{\chi_i}|\}\leq10$: $(d_{\chi_1}, d_{\chi_2}, d_{\chi_3})=\left\{ (1,1,1), (4,4,-5), (10,17,-18) \right\}$. It is important to note the solutions with \textit{unequal} $d_{\chi_i}$.

\begin{figure} 
\includegraphics[width=0.45\textwidth]{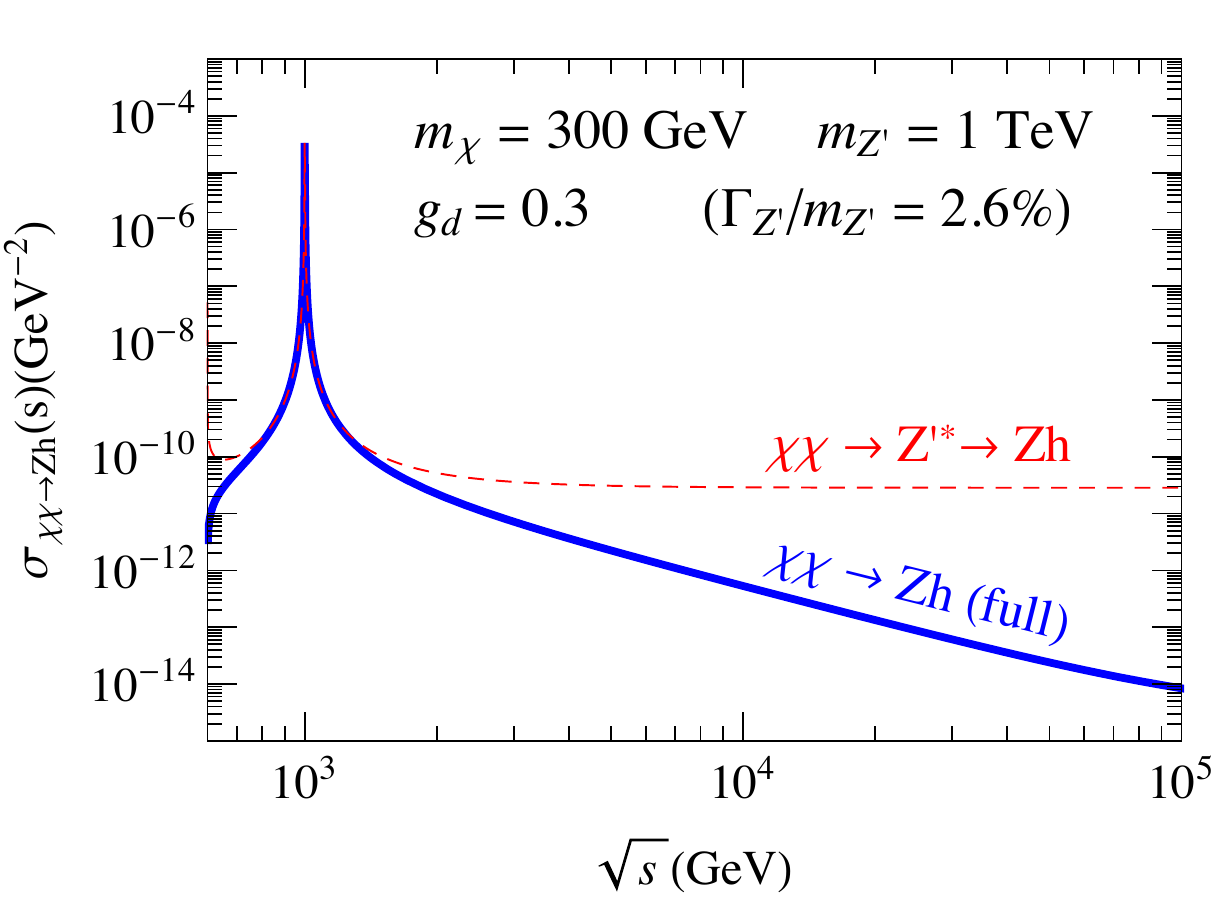}
\caption{Cross section for $\chi \chi \rightarrow Z h$ as a function of $\sqrt{s}$. The  $Z'$ exchange diagram alone (dashed red) signals a breakdown of unitarity at large $\sqrt{s}$. Summing over both $Z$ and $Z'$ exchange diagrams restores unitarity (blue).}
\label{fig:CrossSectionZh}
\end{figure}

For a given solution of \Eq{eq:generalSMsinglets}, we evaluate the quantity $\xi \equiv - (1/3) \sum_{i=1}^n d_{\chi_i}$. This identifies the dark charges of SM fields as:
\be
\begin{split}
& \left\{d_Q, d_u, d_d, d_L, d_e, d_H \right\} =  \\ &
\left\{ - \frac{\xi + \tau}{4}, \tau,  \frac{\xi - \tau}{2}, \frac{3 (\xi + \tau)}{4} ,- \frac{\xi + 3 \tau}{2} , \frac{\xi - 3 \tau}{4} \right\}  \ ,
\end{split}
\label{eq:SMchargesSOLlong}
\ee
up to the freedom of choosing $\tau$. This can be re-expressed as a linear combination $d_{f_{\rm SM}} = (\xi - 3 \tau) Y_{f_{\rm SM}} / 2 - \xi (B-L)_{f_{\rm SM}}$. This was discussed in earlier literature such as \cite{Weinberg:1996kr}. Our parametrization makes two important facts manifest. First, $U(1)_{B-L}$ is the only special case with $d_H=0$. Second, the $U(1)'$ charge of the ``lepton portal'' operator, $LH$, is determined by $d_{\chi_i}$, since $d_L+d_H=\xi$. Unless $d_{\chi}=-\xi$, the renormalizable operator $L H \chi$ that allows fast DM decay is forbidden by $U(1)'$ gauge invariance, even without an extra $\ZZ_2$. 
\vspace{0.2cm}\newline
\noindent 
{\bf Unitarity for $\chi \chi \rightarrow Zh$.} Perturbative unitarity of the $S$-matrix can impose critical constraints. Famously, $WW$ scattering unitarity in the SM provides insights into feasible electroweak symmetry breaking theories and puts an upper bound on the Higgs mass~\cite{Lee:1977yc,Lee:1977eg}. Vector portal models are prone to potential unitarity violation since the longitudinal modes grow with energy. Here we present unitarity constraints on the Majorana DM annihilation $\chi\chi\rightarrow Zh$. Our findings have significant phenomenological implications, since this is potentially the leading process for both relic abundance calculation and indirect detection.
The process is mediated by the operators
\be
\begin{split}
\mathcal{L}_{\chi\chi \rightarrow Zh} = & \, g_d \, d_\chi \, \chi^\dag \overline{\sigma}^\mu \chi \, V_\mu + \\ & 
\frac{\sqrt{2} \, g_w^2}{4} \, v_H \, h \,  \left(Z_\mu^{\rm (SM)}  - 2 \frac{g_d d_H}{g_w} V_\mu \right)^2 \ .
\label{eq:LVZh}
\end{split}
\ee
The interactions with mass eigenstates, $Z$ and $Z'$, can be obtained with \Eq{eq:rotation}. Naively, the $Z'$ exchange dominates for $\theta \ll 1$, as both $\chi$ and $H$ are $U(1)'$-charged. We plot in Fig.~\ref{fig:CrossSectionZh} the cross section as a function of the center of mass energy $\sqrt{s}$. We assign charges as in the benchmark model discussed below, and choose parameters giving $\theta \simeq 10^{-3}$. The red line, accounting for $Z'$ exchange only, approaches a constant value at large energy (signaling a breakdown of unitarity!). Adding the $Z$ exchange diagram (blue line) dramatically alters this behavior. We also notice a closely related significant difference in the non-relativistic limit of $\sqrt{s} \simeq 4 m_\chi^2$, crucial for DM phenomenology as discussed later in this work.
    \newline
  \indent  The $Z$ exchange diagram has to be taken into account, even for very small $\theta$. This is because the $Z'$ diagram also vanishes for $\theta \rightarrow 0$, as the $ZZ'h$ coupling arises from the cross term in \Eq{eq:LVZh} and $\tan2\theta\propto d_H$ (see \Eq{eq:mixingangle}). The two contributions are potentially comparable in size. 
Although there is no manifest explanation for why the two diagrams should destructively interfere as in Fig.~\ref{fig:CrossSectionZh}, such a precise cancellation is critical as it saves the model from the potential violation of S-matrix unitarity, which we explain as follows. \newline
     \indent For a generic $i\rightarrow j$ process, with matrix element decomposed in partial waves $\mathcal{M}_{ij} = 16\pi \sum^\infty_{n=0}(2n+1) a_{n, ij}P_n(\cos\theta)$, unitarity bounds ${\rm Re}(a_{n, ij})\leq1/2$ at large $\sqrt{s}$. The amplitude in this limit is dominated by internal and external longitudinal modes. The $Z'$ exchange diagram results in
\be
\begin{split}
\mathcal{M}_{\chi \chi \rightarrow Z_L h}^{(Z_L^\prime)} =  & \, \sqrt{2} g^2_w g_d d_H \, \frac{v_H m_\chi}{m_{Z'}^2 m_Z} \, c_\theta \, \sqrt{s} \; \times \; \\ &
\left( c_\theta + 2 \frac{g_d d_H}{g_w} s_\theta \right) \left( c_\theta 2 \frac{g_d d_H}{g_w} - s_\theta \right) \ . 
\end{split}
\label{eq:MZhZprime}
\ee
This expression badly violates perturbative unitarity at large $\sqrt{s}$. In order to restore unitarity, there must be additional diagram(s) destructively interfering with \Eq{eq:MZhZprime}, which in this case is the $Z$ exchange diagram
\be
\begin{split}
\mathcal{M}_{\chi \chi \rightarrow Z_L h}^{(Z_L)} =  & \, \sqrt{2} g^2_w g_d d_H \, \frac{v_H m_\chi}{m_Z^3} \, s_\theta \, \sqrt{s} \; \times \; \\ &
\left( c_\theta + 2 \frac{g_d d_H}{g_w} s_\theta \right)^2 \ . 
\end{split}
\label{eq:MZhZ}
\ee
The neat cancellation between the two amplitudes in Eqs.~(\ref{eq:MZhZprime}) and (\ref{eq:MZhZ}) is not yet all obvious at this step, since the sum of the two diagrams is proportional to
\be
\begin{split}
\mathcal{M}_{\chi \chi \rightarrow Z_L h} \propto & \,  \left[  \frac{s_\theta}{m_Z^2} \left( c_\theta + s_\theta 2 \frac{g_d d_H}{g_w}  \right) +  \right. \\ & \left. \;\;\; \frac{c_\theta}{m_{Z^\prime}^2}  \left( c_\theta 2 \frac{g_d d_H}{g_w} - s_\theta \right)  \right]  \sqrt{s} \ .
\end{split}
\label{eq:unitarityfinal}
\ee
Ultimately the cancellation can be made manifest after working out the algebra with \textit{exact expressions} of $m_{Z,Z'}$ and $\theta$ (see \Eq{eq:mixingangle}) in terms of the Lagrangian parameters. After doing so, the expression in \Eq{eq:unitarityfinal} vanishes. 
In addition to explaining the puzzling cancellation between scattering amplitudes, S-matrix unitarity also put bounds on model parameters. We consider the full amplitude for the $\chi \chi \rightarrow Z_L h$ scattering. For internal longitudinal propagators, the subleading term (following the one in \Eq{eq:unitarityfinal}) goes as $1 / \sqrt{s}$ and thus respects unitarity. The amplitude for internal transverse propagators does not diverge at large $\sqrt{s}$, but is dominated by a constant term subject to the unitarity bound. The leading contribution comes from $n=0$ in the partial-wave expansion
\be
a_{0, \, \chi \chi Z_L h}  = \frac{\sqrt{2}}{128} \frac{g_w v_H}{m_Z} \, g^2_d \, d_\chi d_H \, \left( c_\theta + s_\theta 2 \frac{g_d d_H}{g_w}  \right)\ .
\ee
In the small mixing angle limit, justified by EWPT, perturbative unitarity imposes the following bound
\be
g_d \, |d_\chi d_H|^{1/2}  \lesssim 4 \sqrt{2} \;  \ .
\label{eq:unitaritybound}
\ee 
\noindent\textbf{A Benchmark Model.} We explore a benchmark model and its phenomenology, with emphasis on the theoretical issues discussed above. We add three fermion singlets with equal dark charges $d_{\chi_i} = 1$, satisfying \Eq{eq:generalSMsinglets} and giving $\xi = - 1$. Majorana mass terms for dark fermions may originate from their Yukawa interactions with the condensing scalars in the dark sector. The lightest dark fermion $\chi_1 = \chi$ plays the role of DM. By choosing $\tau = 1$, we only couple the mediator to electroweak singlets. The dark charges of SM fields as obtained from \Eq{eq:SMchargesSOLlong} read $\left\{d_Q, d_u, d_d, d_L, d_e, d_H \right\} = \left\{0, 1, -1, 0 ,- 1, -1 \right\}$. As $LH\chi$ is $U(1)'$ invariant with this choice, we need a $\ZZ_2$ to ensure DM stability.

\begin{figure} 
\includegraphics[width=0.45\textwidth]{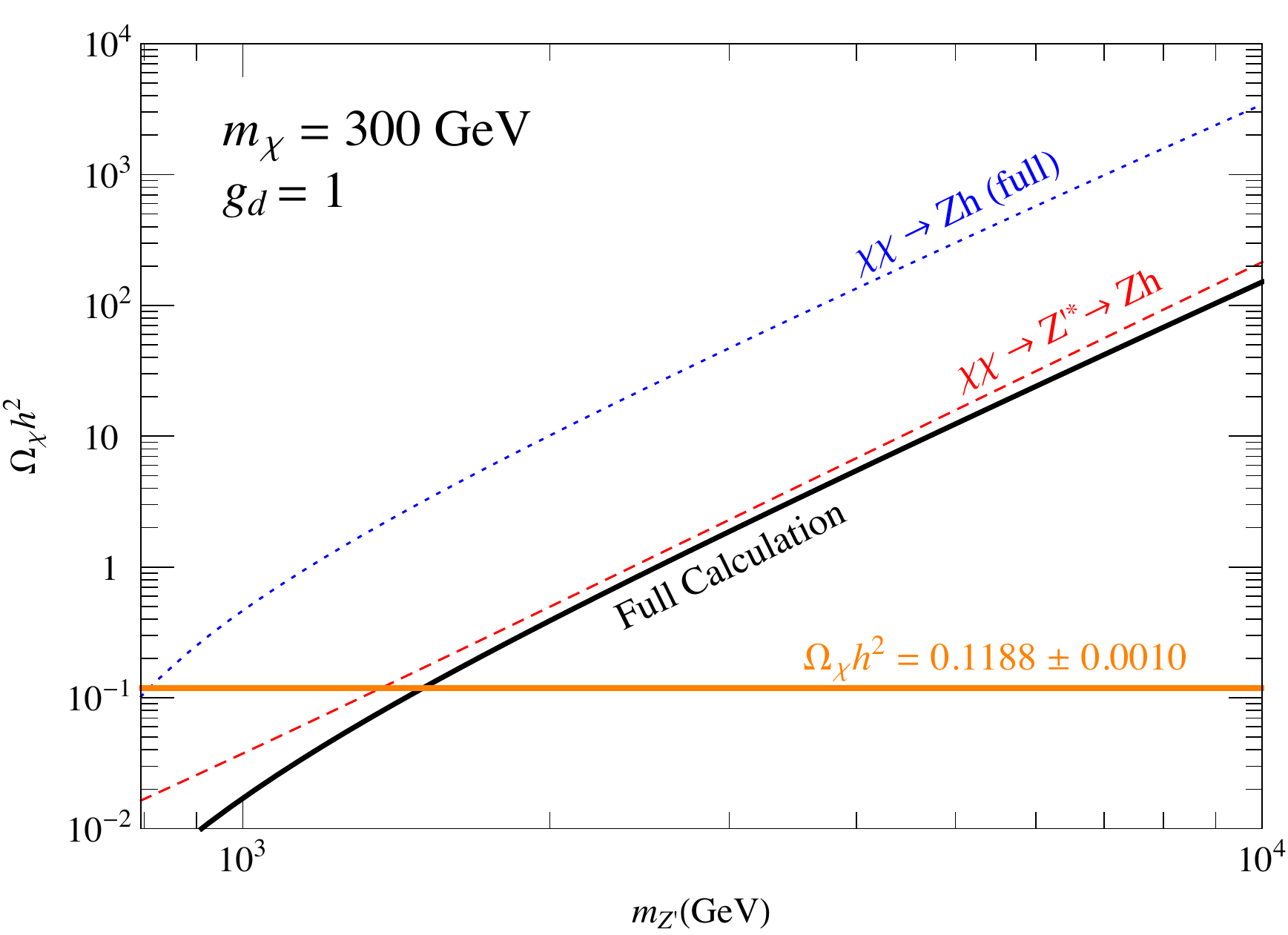}
\caption{DM relic density for the benchmark model as a function of the $Z'$ mass. The full calculation (black) is compared with the result for annihilations to $Zh$, accounting for $Z'$ exchange only (dashed red) and both $Z$ and $Z'$ (dotted blue).}
\label{fig:RelicDensity}
\end{figure}

We focus on the mass region $m_{Z'} > m_\chi \gtrsim 100 \, {\rm GeV}$, and assume that DM can only annihilate to SM final states. Annihilations to SM fermions are p-wave for massive Majorana DM, so the $Zh$ channel could potentially dominate thermal production. This could indeed be the case, as the cross section including only $Z'$ exchange is s-wave. However, as explained earlier, the $Z$ exchange diagram also needs to be included, otherwise we lose unitarity at high energy. The effect is a destructive interference between the two diagrams, leaving the final cross section p-wave suppressed. We show the DM relic density in Fig.~\ref{fig:RelicDensity} for $m_\chi = 300 \, {\rm GeV}$ and $g_d = 1$. Accounting for a virtual $Z'$ only (red line) leads to a much smaller relic density than the sum of the two diagrams (blue line). Therefore DM annihilations to SM fermions dominates, due to the family and color multiplicities, and ultimately set the thermal relic density in our benchmark model. Another consequence of the p-wave nature of DM annihilation to $Z h$ is that indirect searches are ineffective. 

\begin{figure} 
\includegraphics[width=0.45\textwidth]{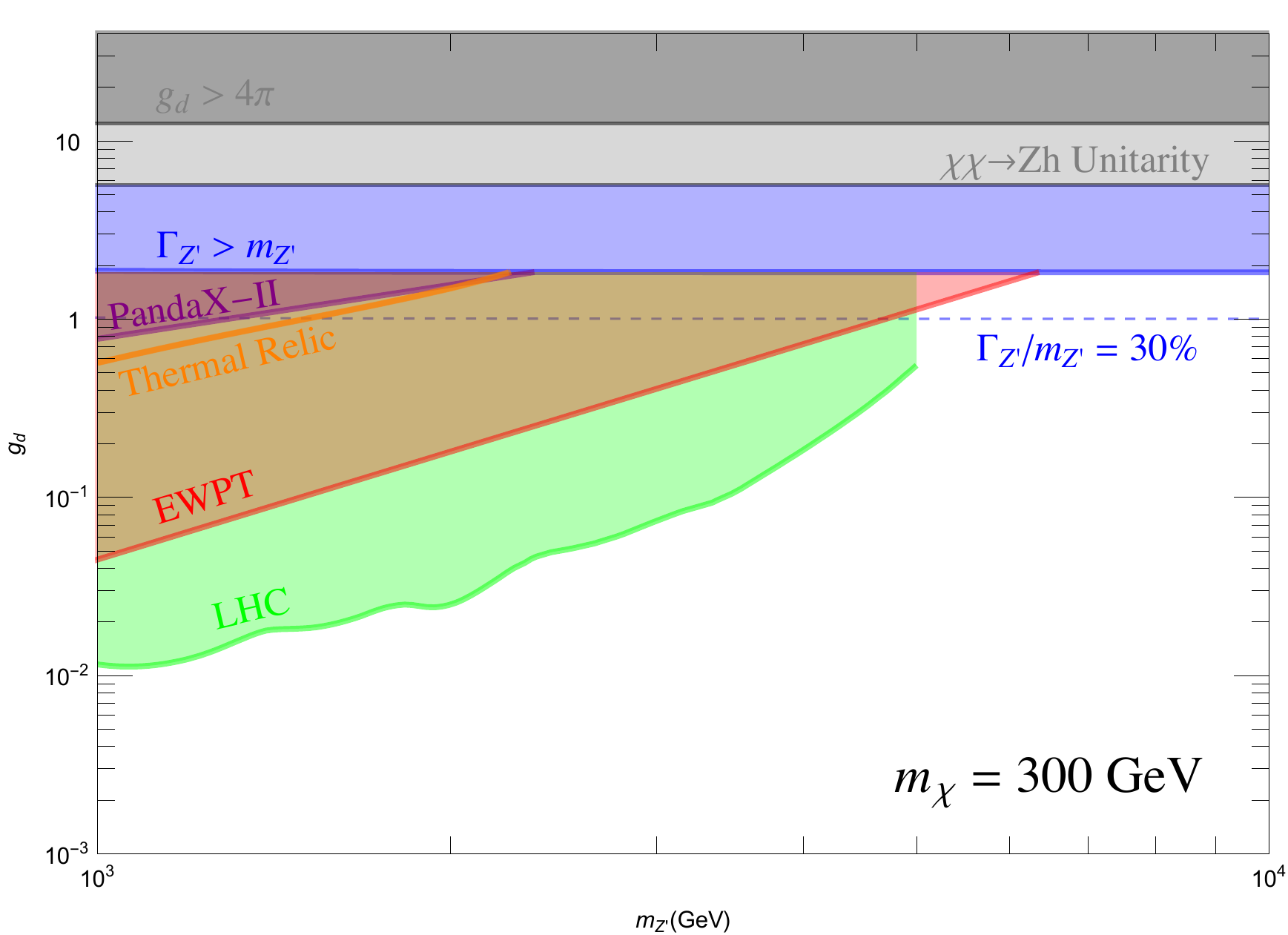}
\caption{Phenomenological constraints for $m_\chi = 300 \, {\rm GeV}$. LHC dilepton bounds~\cite{ATLAS-CONF-2017-027, Khachatryan:2016zqb} are imposed by evaluating the parton distribution functions from Ref.~\cite{Martin:2009iq}. Renormalization group effects for direct detection~\cite{Crivellin:2014qxa,D'Eramo:2014aba,DEramo:2016gos} are accounted for by evolving the couplings with the code \runDM~\cite{runDM}.}
\label{fig:phenoplot}
\end{figure}

The experimental results that do constrain our model are shown in Fig.~\ref{fig:phenoplot}, where we fix $m_\chi = 300 \, {\rm GeV}$ and we identify the allowed region in the $(m_{Z^\prime}, g_d)$ plane. First, we shade away the regions where the gauge coupling is non-perturbative and when the unitarity bound in \Eq{eq:unitaritybound} is violated. Then we also outline the region when the mediator width becomes comparable to its mass. Due to the nearly universal couplings to the SM and dark fields as required by gauge invariance and anomaly cancelation (e.g. \Eq{eq:SMchargesSOLlong}), the $Z'$ search at the LHC naturally involve complementarity among different decay channels. Dilepton searches leads to the strongest constraint, except for large mediator masses ($m_{Z^\prime} \gtrsim 4 \, {\rm TeV}$) where the EWPT bound~\cite{Erler:2009jh} is stronger. Nevertheless, due to the aforementioned intrinsically large multiplicity of $Z'$ decay channels in this model, $Z'$ can be a wide resonance even with perturbative couplings (as shown in Fig.~\ref{fig:phenoplot}). In the plausible case where the $Z'$ width is well over $30\%~m_{Z'}$, the LHC dilepton resonance constraint may not apply and it is then more important to consider complementary channels, in particular $Zh$ \cite{ATLAS-CONF-2017-018, CMS-PAS-B2G-16-007}. Direct detection bounds from PandaX-II~\cite{Fu:2016ega} and PICO~\cite{Amole:2017dex} lead to similar constraints. The region where DM can be produced by standard thermal freeze-out is excluded. Solutions can come from dilution, naturally arising from motivated non-standard cosmologies~\cite{McDonald:1989jd,Kamionkowski:1990ni,Chung:1998rq,Giudice:2000ex,Kane:2015jia,Co:2015pka}, or by enriching the dark sector content to allow additional annihilation channels \cite{DEramo:2010keq, Belanger:2011ww, Agashe:2014yua, Berger:2014sqa, Chacko:2015noa}. Also note that we chose to plot away from resonance enhanced annihilation region which requires some degree of tuning in $m_{\chi}, m_{Z'}$, but may simultaneously accommodate a thermal relic abundance and other constraints.

 \vspace{0.1cm}

\noindent 
{\bf Discussion.} In this paper, we highlighted how a complete understanding of vector portal theories requires a thorough consideration of the following theoretical issues: gauge invariance of SM Yukawa interactions, anomaly cancellation and S-matrix unitarity. We systematically studied the broad class of singlet-only extension models. The phenomenological consequences are considerable, as shown in the benchmark model studied in this work. 

Our analysis opens up several future research directions. For the singlet-only extensions studied here, peculiar models of phenomenological interests such as leptophilic, leptophobic or pure axial-vector coupling to SM quarks are not allowed, as one can see from \Eq{eq:SMchargesSOLlong}. This motivates a systematic study of the realizations of these possibilities, with extra SM charged exotics and/or flavor-dependent $Z'$ couplings, along the lines of the explicit solutions found in Refs.~\cite{Carone:1994aa,Carena:2004xs,FileviezPerez:2010gw,Ko:2010at,Ko:2011ns,Dudas:2013sia,Perez:2014qfa,Dobrescu:2014fca,Hooper:2014fda,Feng:2016ysn,Ismail:2016tod,DelleRose:2017xil,Ellis:2017tkh, Dror:2017nsg,Ismail:2017ulg}.

As shown for one benchmark, the UV-completeness of vector portal theories have profound implication for DM and Z' complementarity. The charges of both SM and dark sector fields are tied to each other, as required by classical and quantum gauge invariance of the Lagrangian. This connects different experimental searches. The anomaly condition cannot be satisfied by only one new fermion, so the dark sector must be richer than just one DM particle. This motivates searches for additional dark sector fermions, whose couplings are predicted by the anomaly condition. Furthermore, the interplay between DM and Higgs physics is natural and essential: unless the new abelian group is proportional to $U(1)_{B-L}$, the SM Higgs doublet is $U(1)'$-charged. This implies the complementary signal in the $Z' \rightarrow Z h$ channel at the LHC~\cite{Aaboud:2016lwx,Khachatryan:2016cfx} with predicable events rates with respect to the dilepton channel. A mediator lighter than the weak scale can gve rise to exotic Higgs decay~\cite{Curtin:2014cca}. 

Finally, although we studied a benchmark with weak scale DM and $Z'$, our results are valid for other DM and mediator masses. In particular, the anomaly condition must be satisfied for a sub-GeV dark photon. The possibility for the SM dark charges in \Eq{eq:SMchargesSOLlong} is different and much more diverse from the one arising from the widely considered abelian kinetic mixing, where they are proportional to the electric charge and thus parity-conserving. This therefore motivates a systematic study of sub-GeV dark sector phenomenology in light of our results. 

\medskip
{\it Acknowledgments.}--- We thank Bogdan Dobrescu, Howard Haber, Bradley Kavanagh, Paul Langacker, Josh Ruderman and Jesse Thaler for helpful discussions. The authors thank the Aspen Center for Physics for hospitality and support when this work was initiated through National Science Foundation Grant No. PHY-1066293. FD was supported by the U.S. Department of Energy grant number DE-SC0010107.

\bibliography{refs_zpdm}

\end{document}